\documentclass[preprint2]{aastex}
\usepackage{graphicx}
\usepackage{pdfpages}
\usepackage{aas_macros}  
\usepackage{amsmath}
\usepackage{chngcntr}

\newcommand{\pilar}{Pilar Monta\~n\'es-Rodr\'{\i}guez }

\slugcomment{}

\shorttitle{Jupiter as an exoplanet}
\shortauthors{\pilar}

\begin{document}


\title{Jupiter as an exoplanet: UV to NIR transmission spectrum reveals hazes, a Na layer and possibly stratospheric H$_2$O-ice clouds}


\author{\pilar\altaffilmark{1}, Gonz\'alez-Merino\altaffilmark{1}, B.; Pall\'e, E.\altaffilmark{1}}

\email{pmr@iac.es}

\affil{Instituto de Astrof\'{\i}sica de Canarias, C/V\'{\i}a L\'actea s/n, E-38200 La Laguna, Spain}

\author{L\'opez-Puertas, Manuel}

\affil{Instituto de Astrof\'{\i}sica de Andaluc\'{\i}a (CSIC), Glorieta de la Astronom\'{\i}a s/n, E-18080 Granada, Spain}

\and

\author{Garc\'{\i}a-Melendo, E.\altaffilmark{2},\altaffilmark{3}}

\affil{Departamento de F\'{\i}sica Aplicada I, E.T.S. Ingenier\'{\i}a, Universidad del Pa\'{\i}s Vasco, 9 Alameda Urquijo s/n, 48013 Bilbao, Spain.}


\altaffiltext{1}{Departamento de Astrof\'{\i}sica, Universidad de La Laguna, Av., Astrof\'{\i}sico Francisco S\'anchez, s/n, E-38206 La Laguna, Spain}
\altaffiltext{2}{Fundaci\'o Observatori Esteve Duran. Avda. Montseny 46, Seva 08553, Spain}
\altaffiltext{3}{Institut de Ci\`encies de l’Espai (CSIC-IEEC), Campus UAB, Facultat de Ci\`encies, Torre C5, parell, 2a pl., E-08193 Bellaterra, Spain}


\begin{abstract}

Currently,  the analysis of transmission spectra is the most successful technique to probe the chemical composition of exoplanet atmospheres. But the accuracy of these measurements is constrained by observational limitations and the diversity of possible atmospheric compositions. Here we show the UV-VIS-IR transmission spectrum of Jupiter, as if it were a transiting exoplanet, obtained by observing one of its satellites, Ganymede, while passing through Jupiter’s shadow – i.e., during a solar eclipse from Ganymede. The spectrum shows strong extinction due to the presence of clouds (aerosols) and haze in the atmosphere, and strong absorption features from  CH$_4$. More interestingly, the comparison with radiative transfer models reveals a spectral signature, which we attribute here to a Jupiter stratospheric layer of crystalline H$_2$O ice.  The atomic transitions of Na are also present. These results are relevant for the modeling and interpretation of giant transiting exoplanets. They also open a new technique to explore the atmospheric composition of the upper layers of Jupiter's atmosphere. 
\end{abstract}





\section{Introduction}

Over the past two decades, more than 1800 exoplanets have been discovered, \citet{5.}, approximately 65{\%} of which are transiting. For a small sample of these planets -- those which orbit bright stars, and have large planet-to-star area ratios -- their atmospheres can be explored through transmission spectroscopy.  During the transit of a planet in front of a star the stellar flux is partially blocked, but a very small fraction of the stellar flux, 10$^{-4}$ for a Jupiter-like planet and a sun-like star system, passes through the thin planetary atmosphere (if an atmospheric thickness of 0.1 times the radius of Jupiter is considered). Transmission spectroscopy has allowed the detection of atmospheric Na I, H I, C II, O I, H$_2$O, CH$_4$, and CO$_2$ \citep{2.,3.,4.,5.,6.,7.}. These observations, however, push the detection capabilities of space and ground-based observatories to the limit and the results are often a source of discrepancies in the literature \citep{6.,7.,8.}, which need to be addressed. Thus, observing planetary transits in our own Solar System can serve as an invaluable benchmark, and provide crucial information for future exoplanet characterizations.

\begin{figure*}
\begin{center}
\includegraphics[width=15cm]{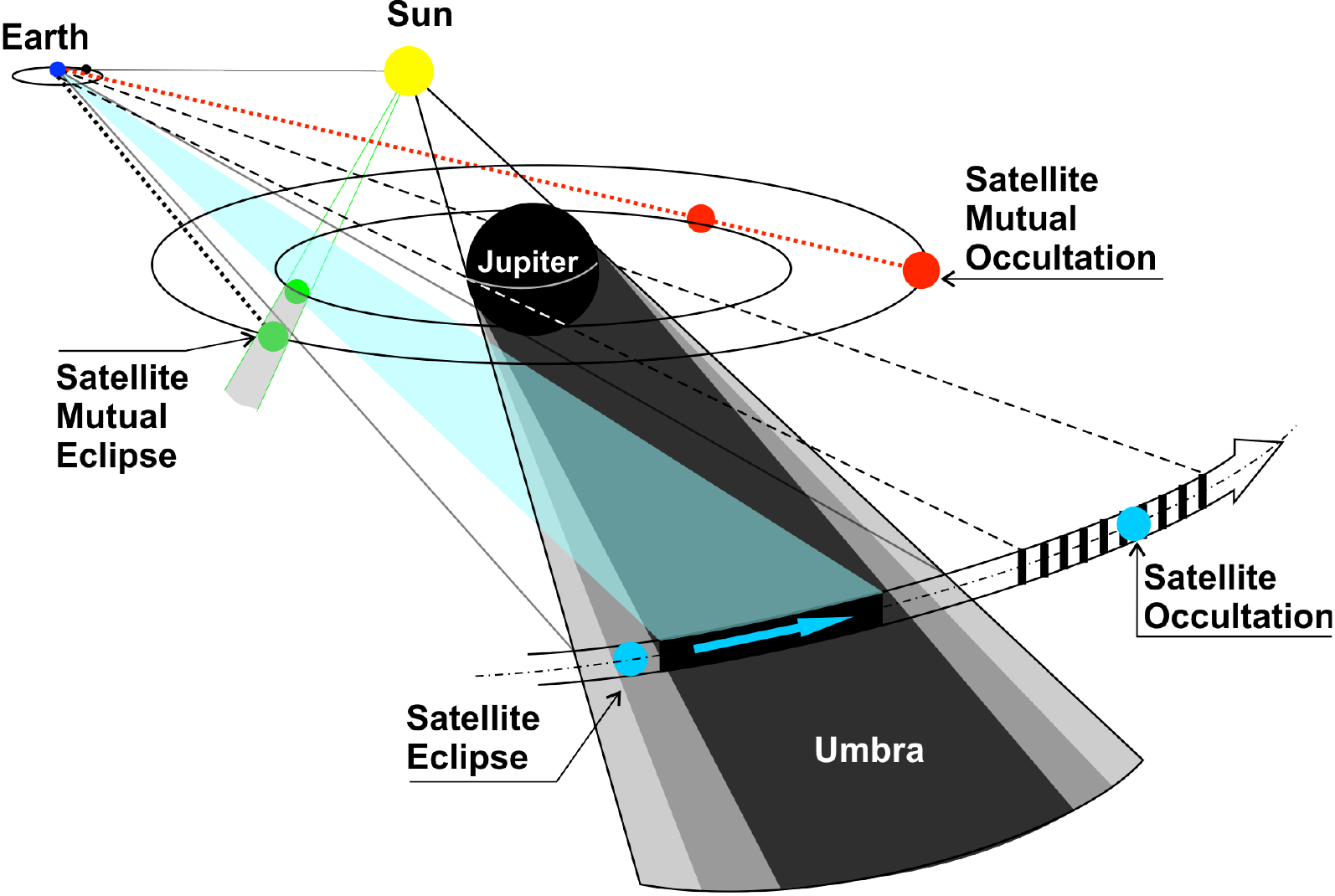}
\includegraphics[width=15cm]{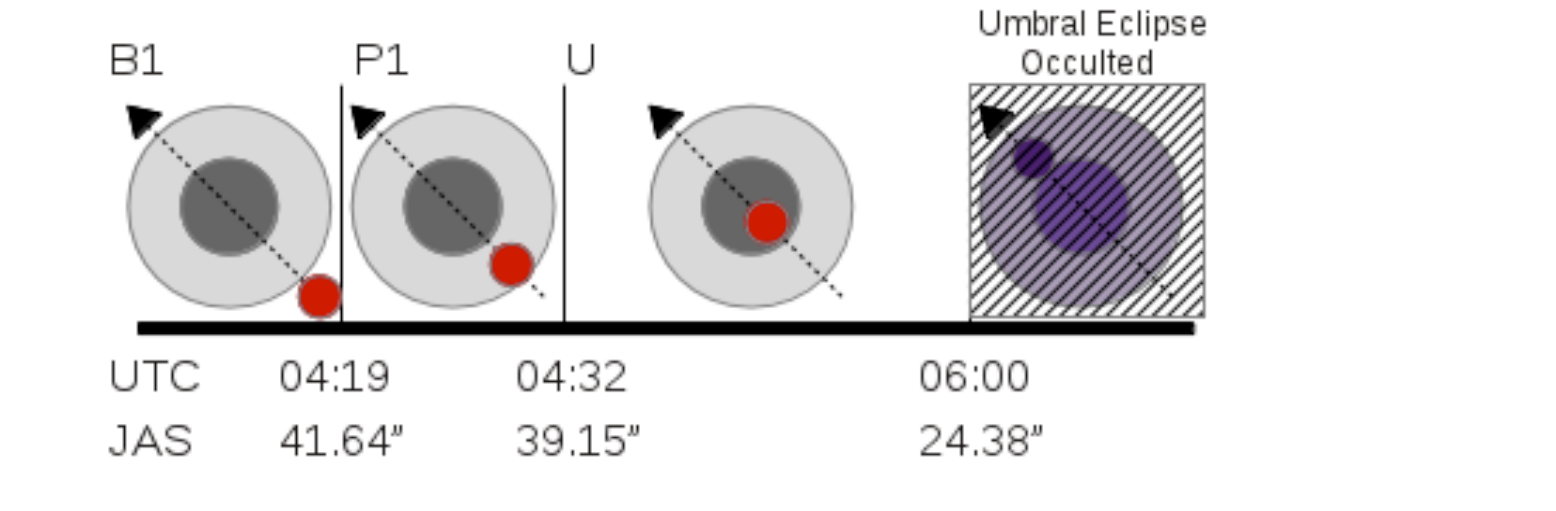}
\caption{Top: A not-to-scale diagram showing the orbital geometry of the Jovian System during the observations. These events are observable from the ground when the Earth is near quadrature. The ratio spectra of Ganymede when in the umbra (or penumbra) and when fully sunlit, results in the determination of the transmission spectrum of Jupiter. Bottom: Universal time (UTC) and Angular Separation from Ganymede center to Jupiter center (JAS) for the eclipse of the satellite on 18/11/2012, as seen from Paranal, Chile.}
\label{fig1_esquema}
\end{center}
\end{figure*}

Here, we report the transmission spectrum of Jupiter, with high signal-to-noise ratio, as if it were a transiting planet. Our technique is to observe Ganymede, which is in synchronous rotation around Jupiter, when crossing Jupiter's shadow. During the eclipse, the spectral features of the Jovian atmosphere are imprinted in the sunlight that, after passing through Jupiter’s planetary limb, is reflected from Ganymede towards the Earth, see Figure~\ref{fig1_esquema}. The ratio spectrum of Ganymede before and during the eclipse removes the spectral features of the Sun, the local telluric atmosphere on top of the telescopes, and the spectral albedo of Ganymede. Ganymede and Europa are practically atmosphere-less bodies and do not introduce any significant variability in the spectra. Similar observations have previously been applied to retrieve the Earth's transmission spectrum, through lunar eclipse observations \citep{9.}.

\section{Observations}

Initially, we observed an eclipse of Ganymede on 06/10/2012, using LIRIS \citep{10.} at WHT in La Palma Observatory, Spain. This experiment was later repeated, and the results confirmed, by observing a second eclipse with XSHOOTER \citep{11.} at VLT in Paranal Observatory, on 18/11/2012. Here, we focus the discussion on the VLT data, due to their higher signal to noise ratio, but a detailed analysis of the WHT observations leads us to virtually identical results.

The larger aperture of VLT allowed us to take rapid measurements - these eclipses are only observable from the ground typically a few of times per year and the suitable observing window last several minutes - with high spectral resolution. At the same time, we extended our measurements into the visible and the near-ultraviolet regions, covering from 300 to 2500 nm in a single exposure. This is possible through a dichroic splitting of the beam in three arms: UV, VIS and near-IR. The 0.5$\,$'', 0.4$\,$'' and 0.4$\,$'' slit widths were used respectively for each range, providing averaged R$\,=\,$9100, 17400 and 11300, respectively. The observing method consisted on taking uninterruptedly spectra of Ganymede in stare mode during its translation through Jupiter's shadow. The telescope active optics was used for guiding at Ganymede's non-sidereal rate in order to keep the target properly centered within the slit width during the dark phases of the event. The apparent magnitude of Ganymede is V=4.6 but it decreases by about 8 magnitudes during the umbra phase \cite{27., 43.}.

Observations started at 3:30 UT, with a seeing below $1''$ which remained stable during the night. The telescope flexure compensation procedure was carried out at 04:00 UT in order to keep the three slits staring at the satellite during the darker phases. The penumbra phase started at 04:19 UT (see Figure~\ref{fig1_esquema}), in which the planet blocks part, but not all, the direct sunlight. During this phase, the amount of direct sunlight reaching the satellite decreases with the progress of the eclipse, whereas the light refracted by the planetary atmosphere reaching the satellite increases. The eclipse was total when the umbra phase started and all direct sunlight was blocked, at 04:40 UT. During the umbra, only refracted light from the planetary atmosphere reaches the satellite. The final out-of-the-eclipse phase was not observed due to the occultation of the satellite by Jupiter's disk, which happened at 06:00 UTC. 

The data reduction of the XSHOOTER was performed using the ESO pipeline Reflex. Each individual exposure had enough signal to noise ratio to be analysed independently. Nevertheless, the three CCDs have different readout times, and we took the VIS exposures as references. We then averaged the several Near-IR spectra taken during a VIS exposure, and the closest UV spectra in time to construct a full spectrum from 300-2500 \micron. This way, 41 individual spectra were determined corresponding to sunlit Ganymede before the eclipse (B1), 8 during the penumbra (P1), and three during the Umbra (U), at all wavelengths simultaneously. 

The penumbra transmission spectra were determined by calculating the ratio between the spectra taken during P1 and the spectra taken just before the eclipse B1. The umbra transmission spectra were determined by calculating the ratio between the spectra taken during U and the spectra taken just before the eclipse B1. Again, these ratios will cancel the telluric contribution of the local atmosphere, the solar spectrum intrinsic in the observations, and the spectral signatures of the satellite (\citet{28.}; \citet{9.}; \citet{20.}), leaving only the contribution from Jupiter's atmosphere.

During these observations, the airmass decreases as the eclipse progresses (from 1.565 in B1 to 1.510 in the deepest U), making it easier to identify residual telluric features, because they would be seen in emission in the (pen-)umbra/bright ratios (see Figure~\ref{fig2_trans_vlt}).

\section{Model Simulations}

Transmission spectra of the occultation of the Sun through Jupiter's atmosphere as observed from Ganymede have been computed, simulating the observations of the eclipse in the penumbra and within the first stages of the umbra. The extension (horizontal) of the Sun's disk as it is setting on Jupiter's horizon has been considered. Because of the strong refraction of Jupiter's atmosphere, this makes it possible to sound Jupiter's limb with moderate vertical resolution (a few tens of km) at the lowest tangent heights.

The transmission spectra have been calculated by using the Karlsruhe Optimized and Precise Radiative Transfer Algorithm (KOPRA; \citet{29.}). KOPRA is a well-tested line-by-line radiative transfer model which offers all the necessary physics for studying this problem.
This code was originally developed for use in Earth's atmosphere and has been recently adapted to the atmospheres of Titan and Mars \citep{30., 31.}. Here, the reference Jupiter's atmosphere, including pressure, temperature and the species abundances were taken from \citet{32.}. We include the major gaseous species, CH$_4$ and H$_2$, and other minor as C$_2$H$_2$, CO, H$_2$O and NH$_3$. The concentration of NH$_3$ (only relevant in the lowest troposphere) was taken from \citet{33.}.
The only species with significant concentration that is not included is ethane (C$_2$H$_6$). The absorption of its strongest band in the region of $2.2-2.5$ $\micron$ would be, however, masked by the stronger CH$_4$ bands in that region. 
The molecular spectroscopic data for all species have been taken from the HITRAN compilation, 2012 edition \citep{34.}. 

Rayleigh scattering by molecular hydrogen and helium 
have also been taken into account by including the Rayleigh optical cross sections provided by \cite{40.} for H$_2$ and by  \cite{39.} for Helium. Ro-vibrational absorption bands resulting from collisions between pairs of H$_2$-H$_2$ and H$_2$-He, the so-called collisions induced absorption or CIA, are also significant in the lower Jupiter atmosphere, where they form a smooth feature mainly in the 2.0-2.5~$\mu$m region. Our simulations include this absorption with absorption coefficients at low temperatures, derived by \cite{36.} for H$_2$-H$_2$ pairs, and by \cite{37.}  and \cite{38.}  for H$_2$-He.

In addition, we also included in the simulations Mie scattering by water ice and aerosols.  The optical properties of the crystalline water ice (real and imaginary part of the refractive index) were taken from \cite{42.} For a temperature of 150 K. The aerosol particles were assumed to have a mean radius of 0.25~$\mu$m, in the range of 0.2-0.5~$\mu$m derived by \cite{16.} for the equatorial particles. The real part of the refractive index of these aerosol particles was taken from \cite{41.} and the imaginary part from \cite{16.}.

\section{Results}

\begin{figure*}
\includegraphics[width=\textwidth]{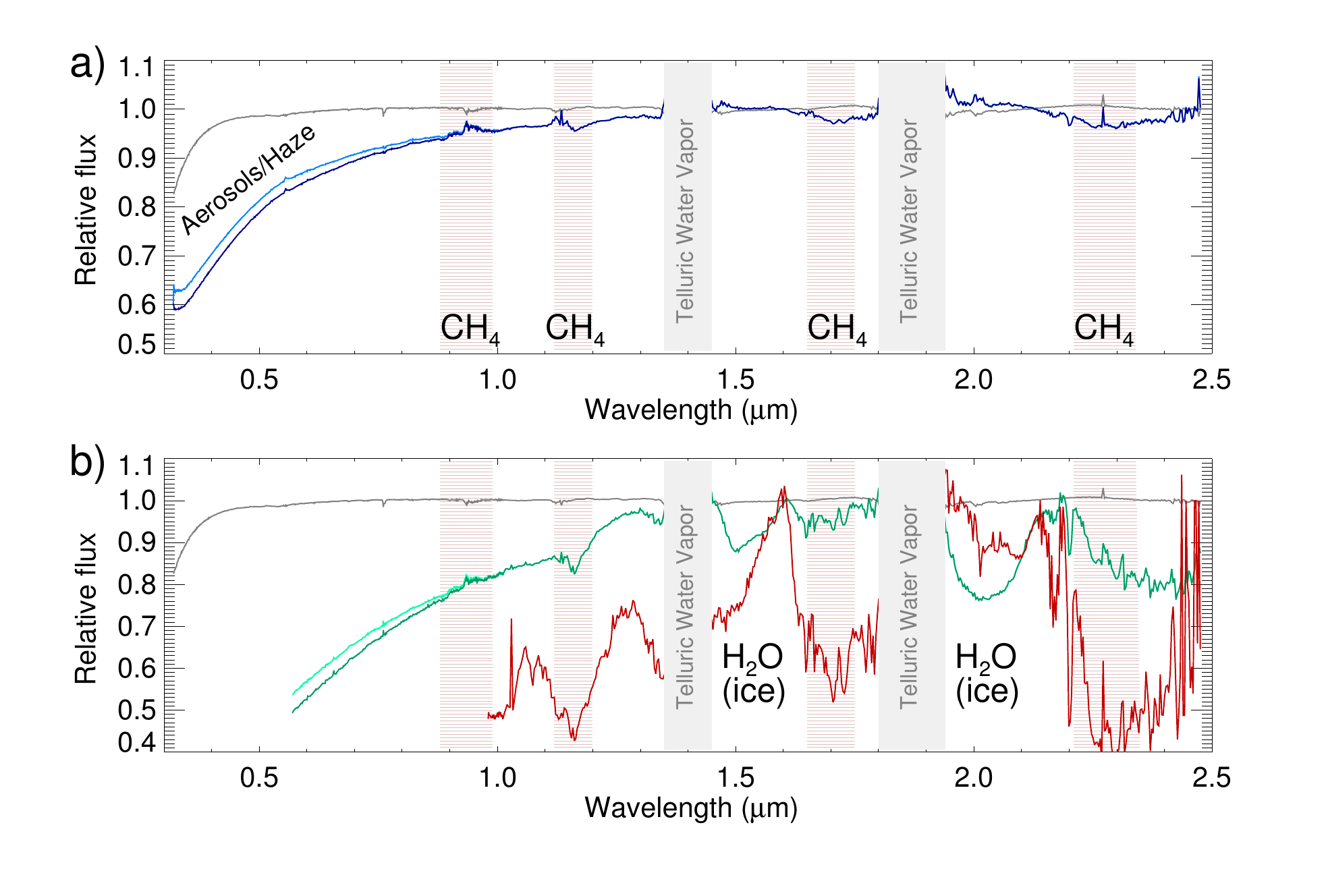}
\caption{The transmission spectrum of Jupiter during the penumbra (panel a) and the umbra (panel b) phases. Gray shaded regions mark the deeply absorbed telluric bands of H$_2$O, which cannot be observed from the ground. The locations of the major $CH_4$ absorption bands are marked with a brown background shadow. On both panels, the thin gray line is the brightness ratio of the light reflected off Ganymede when fully illuminated by the Sun at two different airmass (equivalent to the airmass difference for the two spectra used to extract the transmission spectrum).  It serves to illustrate the contamination from telluric lines to be expected during our measurements. All spectra have been binned to a lower resolution for display purposes. Panel a): Two of our penumbra spectra, taken at different airmass, which are almost identical except for the stronger effect of Jupiter's haze absorption as the eclipse progresses.  Panel b): Three measured umbra spectra, normalized to have unity flux at 1.6 $\micron$. The different relative depth of the bands is due to different timing  within the eclipse umbra, which allows probing different altitudes in Jupiter's upper atmosphere. The first two spectra (green) show features of gaseous $CH_4$ and $H_2O$ ices, while in the third spectrum (red), the $H_2O$ ice features diminish indicating that we are probing below the $H_2O$ ice cloud layer. In the third, deep umbra spectra the spectrum is dominated by the $CH_4$ absorption and by the haze.\label{fig2_trans_vlt}}
\end{figure*}

In Figure~\ref{fig2_trans_vlt}a, the penumbra spectrum of Jupiter is shown, representing a direct transmission component through the upper layers of the atmosphere of the planet. This is directly comparable to the information retrieved through a planetary transit. As expected, the spectra show the major absorption features of the most abundant component, CH$_4$. But most interesting is the change in the spectral continuum: at shorter wavelengths (UV-VIS) transmission is lower due to the extinction caused by aerosols. Extinction by clouds and aerosol particles have a smooth wavelength-dependent extinction effect on the stellar flux which does not produce sharp features. Clouds and hazes have been tentatively detected in Hot Jupiter atmospheres using transmission spectroscopy from the Hubble Space Telescope \citet{6.}, \citet{12.}. Our results give additional confidence to these findings.

In Figure~\ref{fig2_trans_vlt}b, several umbra spectra of Jupiter are shown. As the eclipse progresses, the geometry allow us to sample lower tangent heights, and thus probe lower into Jupiter’s atmosphere. The planetary atmosphere has also the effect of refracting the stellar light \citep{13.}. This refracted component becomes more important as the eclipse progresses deeper into the umbra. Thus, these spectra present a much prominent absorption not only by the aerosol particles (Mie scattering) mainly in the near-IR, but also by the CH$_4$ absorbing bands. However, the most interesting feature is the possible detection of stratospheric H$_2$O ice cloud features at 1.5 and 2.0 \micron.

In Jupiter, the visible cloud deck is composed of NH$_3$ ice condensation, and spans from ~700 up to $200\,mbar$ at the equatorial region, near the tropopause, which is especially conspicuous in planetary images taken in the 890 nm strong methane band \citep{14.}. Below the visible cloud layer, clouds formed by NH$_3$ ice particles, N$H_4$-SH solid particles, H$_2$O ice and H$_2$ONH$_3$ liquid solution are also present, but they are hard to detect \citep{15.}. Above these levels, the major atmospheric constituent is CH$_4$, but recent analysis of observations made with the Cassini Imaging Science Subsystem (ISS), indicates that the planet is wrapped up by a haze at the lower stratosphere at the equatorial region and mid latitudes, at $\approx 50 mbar$, which rises at higher levels (p$\,<\,$20$\,mbar$) over the poles \citep{16.}. But the composition of this layer remained undetermined.

It is from these upper levels that our spectroscopic signals are coming from. Over the umbra evolution, H$_2$O ice absorption features at 1.5 and 2.0 $\micron$ appear and later diminish (see Figure~\ref{fig2_trans_vlt}b) as the sounded tangent height on the planetary limb crosses the altitude (or pressure levels) where these particles are located. Their detection in our transmission spectra allows us to determine that Jupiter's upper atmospheric hazes near the equator are at least partially composed of a cloud deck of H$_2$O ice crystals. While these H$_2$O ice absorption bands are well known, the spectral measurements of VIMS on board of the Cassini spacecraft taken in 2000 did not detect them \citep{19.}.

\begin{figure*}
\begin{center}
\vspace{-5em}
\includegraphics[width=12cm]{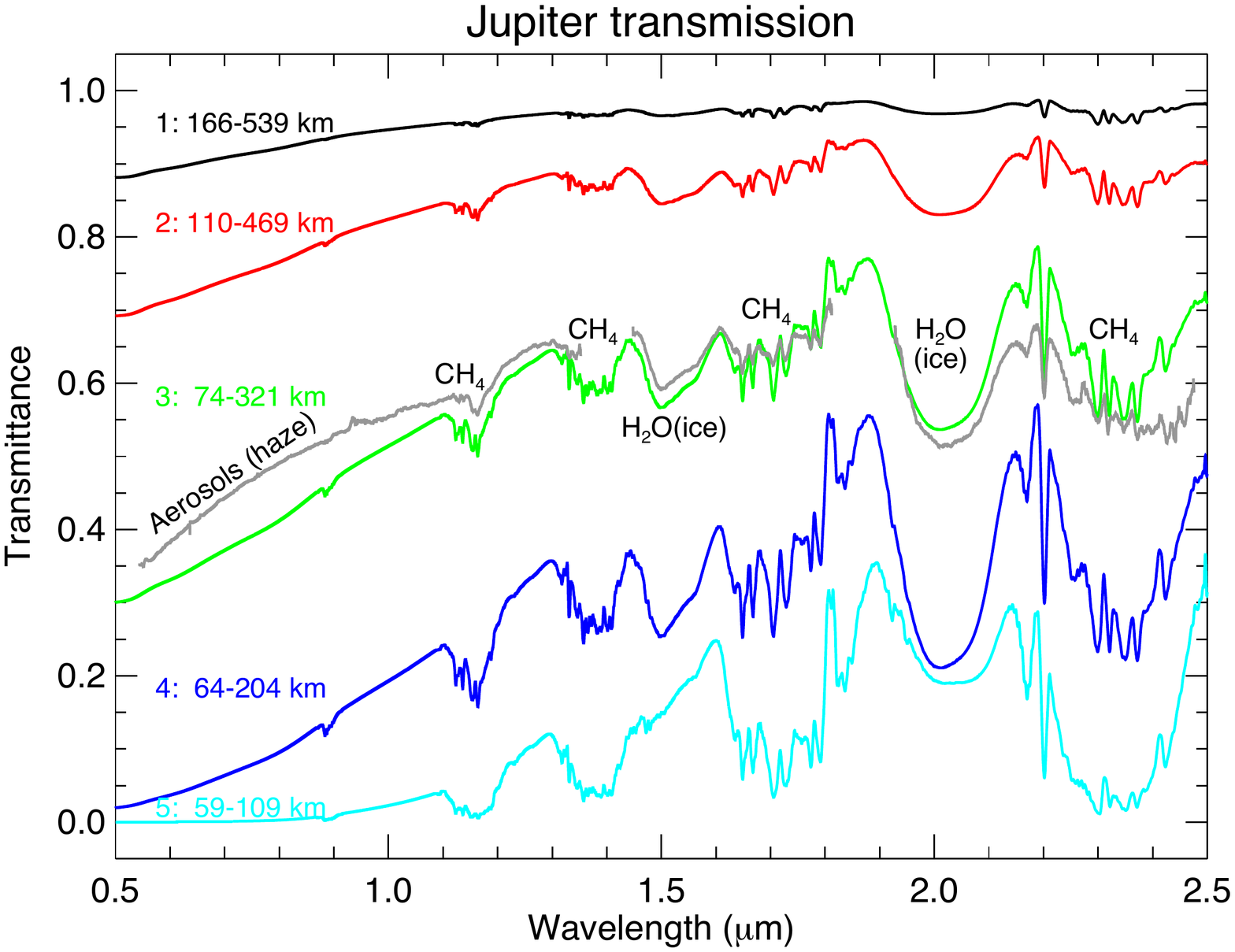}
\includegraphics[width=12cm]{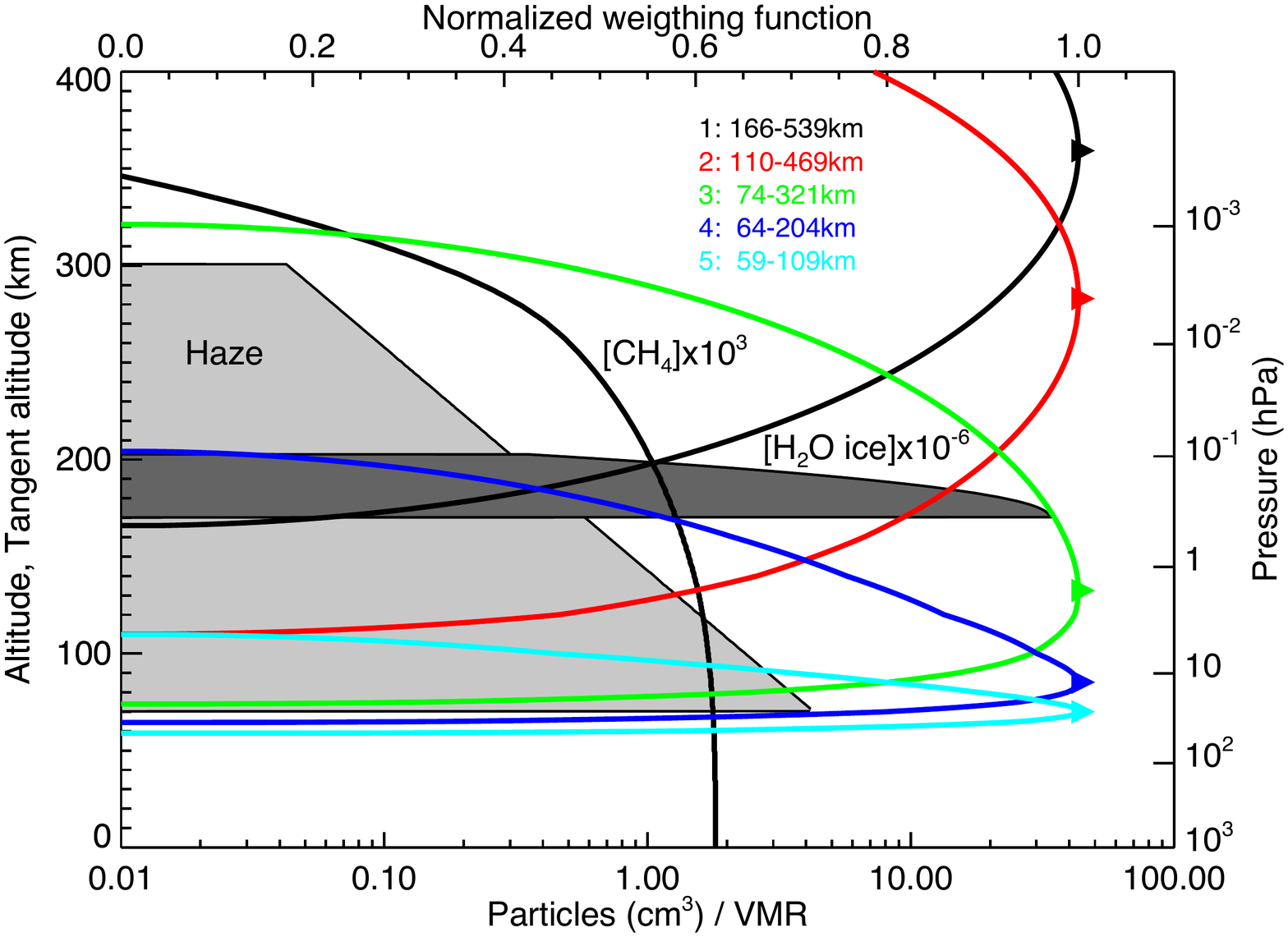}
\caption{Top) Transmission spectra of Jupiter calculated for the early phase of the penumbra, and at several stages during the umbra over the $0.5-2.5$ $\micron$ spectral region. The spectra show the most prominent $CH_4$ bands, the extinction of the aerosol particles (haze), and the two distinct absorption peaks of water ice at 1.5 and 2.0 $\micron$. Plotted with a grey line is one of our observed penumbral spectra from Figure~\ref{fig2_trans_vlt}b.
Bottom) An illustration of the distribution of haze, $CH_4$, and $H_2O$ ice in our model, together with geometrical information on the range of tangent altitudes covered in each of our simulations in the top panel, corresponding to the mean times when the (umbra)-penumbral spectra were taken. Altitude is taken as zero at the 1 bar pressure level. In addition, we have over-plotted the profiles of the concentrations of aerosols and of the water ice particles cloud used in the computation of the transmission spectra. Note how the sounded region shrinks as the eclipse progresses, while moving down to lower tangent heights. 
\label{models}}
\end{center}
\end{figure*}

In Figure~\ref{models}a our model simulations of Jupiter's umbra and penumbra transmission spectra are plotted. All major features of the measured spectra are simulated. The absorption spectral features at 1.5 and 2.0$\,\micron$ can be very well reproduced in our model with crystalline water ice, needing a total column of about $10^{13} particles/cm^2$ with a size of ~0.01$\,\micron$ located near the $0.5 mbar$ level, a water amount 420 times larger than that measured by HERSCHEL, \cite{25.}, which found a water mass load of 1.5$\times 10^{-7} g/cm^{2}$ in the gas phase. While the good spectral match supports the identification of H$_{2}$O ice, the large amounts and the location at low pressure implied by the data are puzzling. Moreover, the presence of water ice at 0.5 mbar does not have an obvious explanation. 

In terms of aerosol mass loading, \cite{16.} derived ~1$\times 10^{-6} g/cm^{2}$ at low latitudes and ~1$\times 10^{-4} g/cm^{2}$ at high latitudes. The H$_2$O ice particle distribution we derive (log normal with sigma=0.3 and r=0.01$\,\micron$) gives a H$_2$O ice mass load of 6.3$\times 10^{-5} g/cm^{2}$. Hence, although it is larger that their aerosol mass load for low latitudes, it is comparable to their mass load for high latitudes, using observations at 0.25 and 0.9 $\,\micron$. Hence, we do not have any conflict between the derived H$_2$O ice load and the results from previous literature.
The haze parameters used in our model are in agreement to those recently reported for Jupiter’s equator by \citet{16.}. 

The solar atomic lines should disappear in the ratio between the (pen- )umbra and bright spectra of  Ganymede. However, some residual features remain, as several effects will alter the shape of the spectral lines, including: i) small Doppler shifts due to the Sun-Ganymede-Earth relative speeds, ii) changes in the solar region contributing to the spectra, ii) instrumental flexures, and iv) Raman scattering producing a center-to-limb brightening \citet{20.}. This latter effect seems to dominate the spectral shape of the ratio spectra near atomic solar lines, where the line residuals have an inverted W-shape pattern.

\begin{figure*}
\epsscale{.80}
\includegraphics{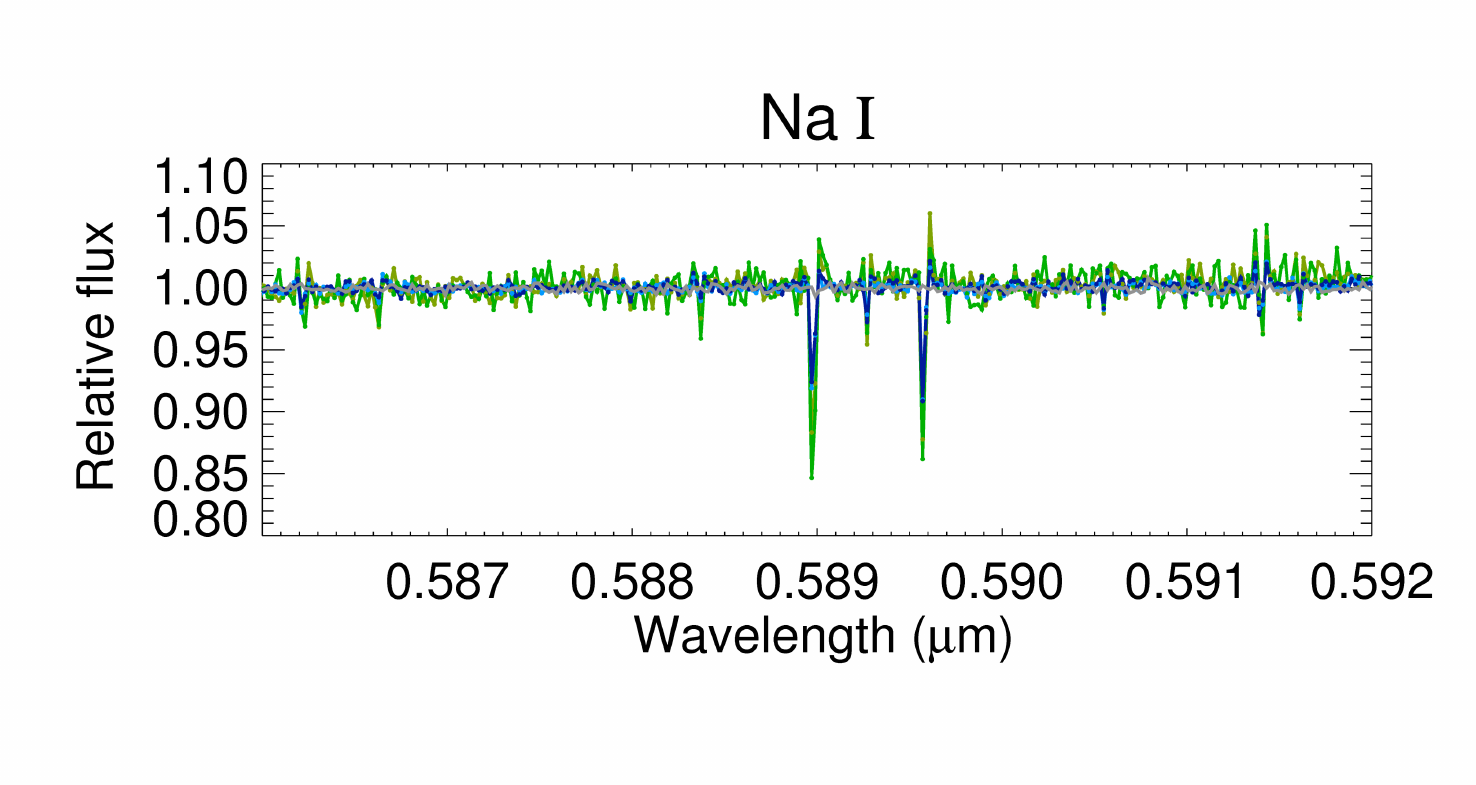}
\caption{The evolution of Ganymede spectra during the observations centered on the Na I doublet wavelength region. The grey spectra are fully illuminated Ganymede spectra immediately before the penumbra phase, divided by a previous spectra also in full sunlight illumination (arbitary reference spectrum). These spectra are basically featureless. In blue are the penumbra spectra divided by the same reference spectrum, and they start to show signs of extra absorption at 0.5889 and 0.5896 $\micron$. In green are the two umbra spectra for which the SNR is good in this wavelength region.}
\label{sodium}
\end{figure*}

In Figure~\ref{sodium} we plot several penumbra and umbra spectra (both divided by the direct sunlit Ganymedes spectrum) focusing on the region near the Na I doublet. Most of the lines are strongly deformed in a classical “W” pattern, as is the case of the Fe lines in between the Na I doublet, and other solar lines at 586.2 and 591.4 $\micron$. 
But in the case of the Na I, there is a net absorption indicating the presence of a Na layer in Jupiter's upper atmosphere.  This Na I absorption starts to be clearly detectable in the last penumbra spectra (at high SNR) and in all umbra spectra (with progressively decreasing SNR). We plan to carry out a detailed study of these features in a future paper. 

The presence of Na in Jupiter's upper atmosphere can be explained by the deposition of either cometary impacts \citep{21.} or the continuous outward flux of Na from Io \citep{22.}. The fact that our temporal series of sunlit Ganymede spectra do not show any signs of Na absorption, and neither do the first penumbra spectra, indicates that the origin of this absorption is located within Jupiter's atmosphere, and not in the torus of Na trailing along Io's orbit. 

\begin{figure*}
\includegraphics[width=\textwidth]{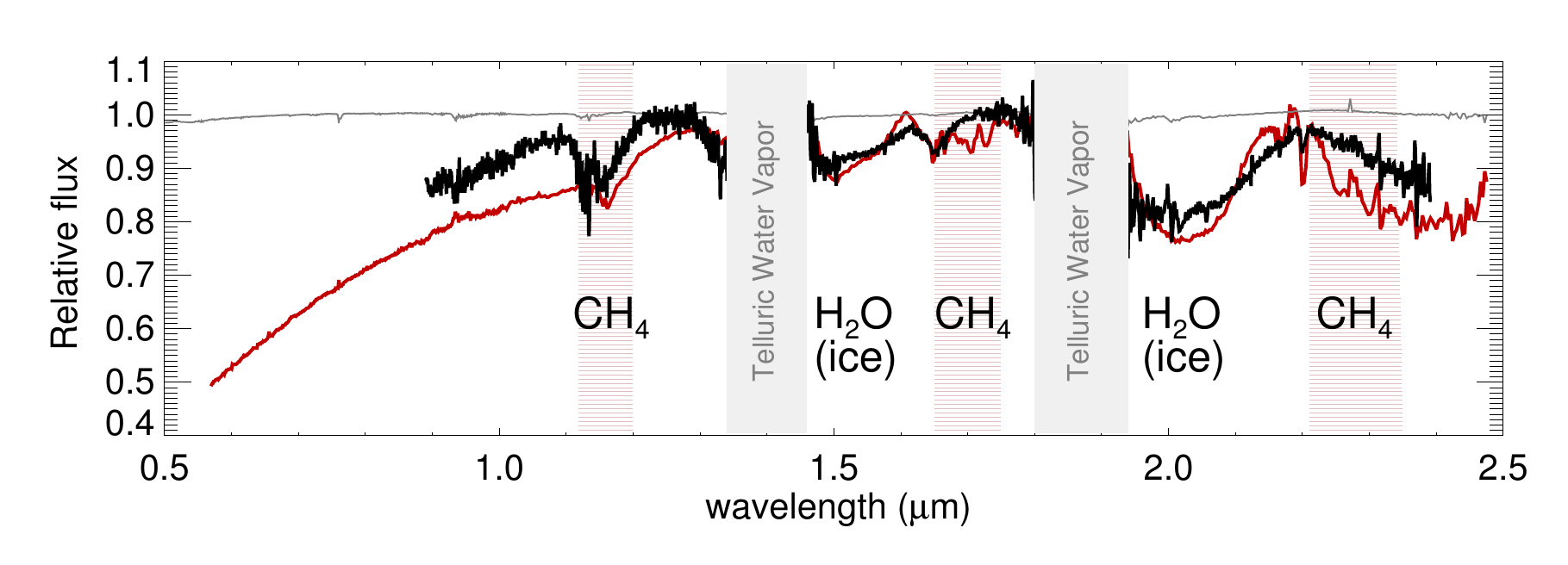}
\caption{ The transmission spectrum of Jupiter as measured with the WHT telescope (black line) during the  umbral phase. Gray shaded regions mark the deeply absorbed telluric bands of H$_2$O, which cannot be observed from the ground. As in Figure~\ref{fig2_trans_vlt}, the thin gray line is the brightness ratio of the light reflected off Ganymede when fully illuminated by the Sun at two different airmass. Overplotted (red line) is one of Jupiter's umbral spectra taken with the VLT. Note how both WHT and VLT spectra shown the fingerprints of gaseous $CH_4$ and $H_2O$ ices.}
\label{whtspec}
\end{figure*}

The WHT observations have not been discussed at lenght in this manuscript, however the results of that earlier campaign are similar to those obtained with VLT. In Figure~\ref{whtspec} the transimission spectrum of Jupiter in one of the umbras of the WHT data is compared to one of the umbra obtained with the VLT. While there are differences in the spectra due to the the fact that the eclipse geometry is not exactly the same, and that WHT needs two different exposures (different times, different umbra depths) to cover the 0.9 to 2.5 $\micron$ range, they show essentially the same spectral signatures.

\section{Conclusions}

In summary, we have determined here the UV-VIS-IR transmission spectrum of Jupiter, as if it were a transiting exoplanet. This transmission spectrum reveals the imprints of strong extinction due to the presence of clouds (aerosols) and hazes in Jupiter's atmosphere, and strong absorption features from  CH$_4$. More interestingly, the comparison with radiative transfer models reveals a spectral signature, which is attributed here to a stratospheric layer of crystaline H$_2$O ice.  
The atomic transitions of Na are also present. These results are relevant for the modeling and interpretation of giant transiting exoplanets, but they also open a new technique to characterize the upper layers of Jupiter's atmosphere. Taking advantage of the scanning of faint features that this technique provides, observations of other satellite eclipses from space could help to set limits to the stratospheric water abundance in the upper layers of Jupiter's atmosphere and provide a way to monitor the rate of cometary impacts on Jupiter \citep{23.} which, in turn, has consequences for the formation history of the solar system.

\acknowledgments

Based on observations made with ESO Telescopes at Paranal Observatory (under program 090.C-0829(A)), and  with the WTH in La Palma. This work is partly financed by the Spanish MINECO through projects 
SEV-2011-0187 (2011 Severo Ochoa Program), AYA2012-39612-C03-02, AYA2010-21308-C03-02, AYA2011-30147-C03-03, AYA2012-36666, and AYA2011-23552. The authors wish to thank M. Hopfner for very valuable discussions about transmission calculations of particles and refraction with KOPRA, L. Lara for supplying the Jupiter reference atmosphere, Agust\'in S\'anchez-LaVega for general discussion, R. Escribano, V. Herrero and B. Mate for valuable discussion on water ice, Emmanuel Lellouch for a very constructive referee review and the ESO and ING support astronomer teams.








\clearpage



\clearpage

\end{document}